\documentclass[twocolumn]{aastex701}
\pdfoutput=1
\usepackage{mathtools}
\usepackage{appendix}
\usepackage{bm}
\usepackage{soul}
\usepackage{natbib}
\usepackage{orcidlink}
\usepackage{newtxtext,newtxmath}
\usepackage[T1]{fontenc}
\usepackage{chemformula}
\usepackage{xcolor}
\usepackage{graphicx}
\usepackage[utf8]{inputenc}
\usepackage{amsmath}
\usepackage{amssymb}
\usepackage{rotating}
\usepackage{fancyhdr}
\usepackage{listings}
\usepackage{placeins}
\usepackage{gensymb}
\usepackage{comment}
\usepackage{array}
\usepackage{tabularx}
\usepackage{url}

\newcommand{\subtractions}[1]{\iffalse #1 \fi}

\newcommand{\Nbin}{\ensuremath{N_{\rm bin}}}

\newcommand{\FAPsig}{$\sigma_{FAP}$}
\newcommand{\psid}{\texttt{psid}}

\newcommand{\tap}[1]{\color{red}#1\color{black}}
\newcommand{\ajd}[1]{\color{orange}#1\color{black}}

\shorttitle{ZVAR DR1 Overview}
\shortauthors{Graham et al.}

\begin{document}

\title{The ZTF Variability Archive (ZVAR): Initial Data Release (DR1)}

\author[0000-0002-3168-0139]{Matthew J. Graham}
\affiliation{Division of Physics, Mathematics, and Astronomy, California Institute of Technology, Pasadena, CA 91125, USA}
\email{mjg@caltech.edu}

\author[0000-0002-6836-181X]{Samuel Whitebook}
\affiliation{Division of Physics, Mathematics, and Astronomy, California Institute of Technology, Pasadena, CA 91125, USA}
\email{sewhitebook@astro.caltech.edu}

\author[0000-0003-0228-6594]{Andrew J. Drake}
\affiliation{Division of Physics, Mathematics, and Astronomy, California Institute of Technology, Pasadena, CA 91125, USA}
\email{ajd@astro.caltech.edu}

\author[0000-0003-2630-8073]{Thomas Prince}
\affiliation{Division of Physics, Mathematics, and Astronomy, California Institute of Technology, Pasadena, CA 91125, USA}
\email{prince@caltech.edu}

\author[0009-0008-3603-0013]{Thomas Culino}
\affiliation{Division of Physics, Mathematics, and Astronomy, California Institute of Technology, Pasadena, CA 91125, USA}
\email{tculino@caltech.edu}

\author[0000-0002-4770-5388]{Ilaria Caiazzo}
\affiliation{Institute of Science and Technology Austria, Austria}
\email{Ilaria.Caiazzo@ist.ac.at}

\author[0000-0002-2626-2872]{Jan van Roestel}
\affiliation{Institute of Science and Technology Austria, Austria}
\email{jcjvanroestel.astro@gmail.com}

\author[0000-0002-0853-3464]{Zach Vanderbosch}
\affiliation{Hobby-Eberly Telescope, McDonald Observatory}
\email{}

\author[0009-0003-6181-4526]{Theo du Laz}
\affiliation{Division of Physics, Mathematics, and Astronomy, California Institute of Technology, Pasadena, CA 91125, USA}
\email{theophile.dulaz@gmail.com}

\author[0000-0002-8262-2924]{Michael W. Coughlin}
\affiliation{School of Physics and Astronomy, University of Minnesota, Minneapolis, MN 55455, USA}
\email{cough052@umn.edu}

\author[0000-0001-7016-1692]{Przemek Mroz}
\affiliation{University of Warsaw, Poland}
\email{pmroz@astrouw.edu.pl}

\author[0000-0002-72226-836X]{Kevin Burdge}
\affiliation{Massachusetts Institute of Technology}
\email{kburdge@mit.edu}

\collaboration{all}{The ZVAR Collaboration}

\correspondingauthor{Matthew Graham}
\email{mjg@caltech.edu}

\begin{abstract}
    The Zwicky Transient Facility (ZTF) has observed the northern sky observable from Palomar Observatory since March 2018, accumulating a median of a thousand photometric measurements across $gri$ for each of more than a billion sources. Among these billions of objects are millions of periodically variable stars, active galactic nuclei, stellar remnants, and substellar objects. 
    We describe the initial data release (DR1) of the ZTF Variability Archive (ZVAR) which provides image difference forced photometry, candidate periods, and
    variability metrics for approximately 1.3 billion sources in ZTF having an identification in Pan-STARRS1 DR1 (PS1).
    The data cover the 5.37 year interval between March 2018 and July 2023. 
    ZVAR is sensitive to periods between 5 minutes and 2.7 years.
    We also describe tools for accessing the ZVAR data.
\end{abstract}

\keywords{
\uat{Time domain astronomy}{2109} ---
\uat{Sky surveys}{1464} --- \uat{Celestial objects catalogs}{212} --- \uat{Time series analysis}{1916}
--- \uat{Period search}{1955} ---\uat{Periodic variable stars}{1213}
}

\section{Introduction} \label{introduction}
\setcounter{footnote}{0}
Recent years have seen the advent of many time-domain focused optical sky surveys geared to the real-time capture of transient astrophysical phenomena, e.g., the Catalina Real-time Transient Survey \citep[CRTS;][]{Drake_CRTS}, the Palomar Transient Factory \citep[PTF;][]{Law_2009}, the All-Sky Automated Survey for Supernova \citep[ASASSN;][]{Kaiser_2004}, and the Asteroid Terrestrial-Impact Last Alert System \citep[ATLAS;][]{Tonry_2018}. The legacy of these, however, is tens to thousands of observations of different regions of the sky at varying cadences in different passbands and to varying depths but over multi-year baselines. These data sets are invaluable for studying persistent variable objects, both systematically at the population level and individually at the object of interest level. Such studies can be further enhanced by maximizing source detections beyond what is commonly provided in data releases through on-request forced photometry services, i.e., measuring the flux at a user-specified position, e.g., ATLAS\citep{Shingles_2021}, and ASAS-SN \citep{Hart_2023}. These services tend to be limited in scope, though, to a few hundred positions in a single query with results delivered asynchronously. 

The Zwicky Transient Facility \citep[ZTF;][]{Bellm_ZTF_overview,Graham_ZTF_objectives,Masci_ZTF_data,Dekany_ZTF_inst} has been observing the visible sky above Dec = $-30^\circ$ from Palomar Observatory in $g$, $r$, and $i$ filters every night since March 19, 2018.
Its camera has a 47 deg$^2$ field-of-view so the entire sky can be covered on a two-to-three night cadence with 30s exposures down to a nominal $5\sigma$ survey depth of 20.6 in the $r$ passband. A real time alert stream reports between 500,000 and 1 million transient detections per night and regular data releases provide light curves for $\sim$1.3 billion sources with more than 20 observations\footnote{https://irsa.ipac.caltech.edu/data/ZTF/docs/releases/ztf\_release\_notes\_latest} (in $r$), although the median number of data points in a light curve is in the hundreds for each filter band.

Machine learning has been applied to ZTF photometry from the data releases to identify variable stars \citep{Chen_ZTFVariables_2020,Sanchez-Saez_ZTFClassifier}, including the ZTF Source Classification Project \citep[SCoPe;][]{Van_Roestel_SCOPE,Coughlin_SCOPE,Healy_SCOPE}, active galactic nuclei \citep[AGN;][]{Nackoneczny_QZO}, cataclysmic variables \citep[][]{Szkody_2020,Szkody_2021}, and young stellar objects \citep{Zhang_YSOs} amongst others. More targeted searches have focused on compact binary objects, including discoveries of the first mass transferring system with a sdOB donor \citep{Kupfer_2020}, a 6.9 minute eclipsing detached double white dwarf system \citep{Burdge_2019}, and the smallest and most massive white dwarfs \citep{Caiazzo_2021}.

The ZTF Forced Photometry Service \citep{Masci_FP} has returned light curves for over 5 million sky positions since 2020. The majority of these requests have been for transient source positions but the service is not designed for large systematic searches over thousands or millions of light curves. In this paper, we introduce the ZTF Variability Archive (ZVAR), a new community repository of ZTF forced photometry light curves and value-added catalogs. In our first data release (DR1), we are making available forced photometry light curves from ZTF difference images at every Pan-STARRS1 source position \citep{chambersPanSTARRS1, Flewelling_PANSTARRS} within the ZTF footprint. Future data releases will feature forced photometry derived from science images. We are also releasing periodicity and variability metrics for ZVAR sources as value-added catalogs and search interfaces on these that support the discovery of new objects on the basis of their variability and periodicity.

This paper is organized as follows: Section~\ref{sec:catalog} describes ZVAR DR1 and its components and Section \ref{sec:data_avail} details the tools and interfaces we are providing to query and access these. We present some illustrative first science results in Section \ref{sec:results} and a discussion of issues connected to the first ZVAR data release in Section \ref{sec:discussion}.
Details of the ZVAR DR1 documentation as well
as file formats of the catalogs are given in the appendices.

\section{The First ZVAR Data Release (DR1)} 
\label{sec:catalog}

\begin{figure*}
    \centering
    \includegraphics[width=\linewidth]{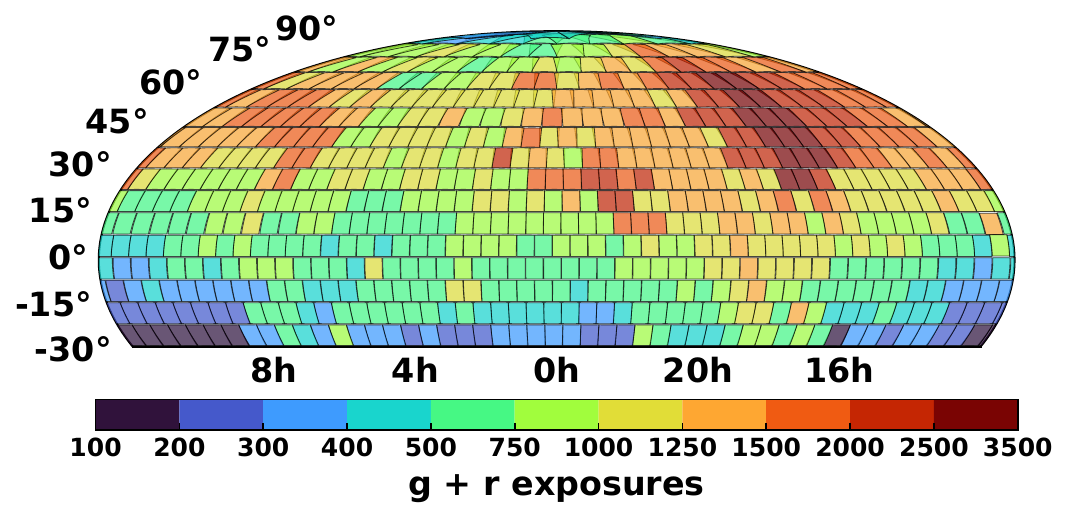}
    \caption{ZVAR coverage in equatorial coordinates split into fields and colored by the number of $g$- and $r$-band forced photometry for ZVAR DR1. The most visited field has 3412 total $g$- and $r$-band exposures.}
    \label{fig:coverage}
\end{figure*}

ZVAR DR1 includes 5 catalogs.  These are summarized here and described
in more detail in the following subsections.
\begin{itemize}
    \item[1.] {\bf Source Catalog}: Lists all sources
    with a PS1 detection within the ZTF footprint for which forced photometry has been performed in ZTF data. The Source Catalog includes PS1 IDs, PS1 mean magnitudes, and Gaia cross-match information as well as the number of observations per filter band.
    \item[2.] {\bf Forced Photometry Catalog}: Contains forced photometry for objects in the ZTF footprint having a PS1 ID.
    Forced photometry is performed on \mbox{ZTF-$g$}, \mbox{ZTF-$r$}, and \mbox{ZTF-$i$} bands
    \citep{Masci_ZTF_data,Dekany_ZTF_inst}. 
    \item[3-4.] {\bf Short Period Catalog (SPC)} and {\bf Long Period Catalog (LPC)}: 
    The SPC covers periods, $P$, in the range
    ($5\ {\rm min} < P < 10\ {\rm days}$) and the LPC covers
    ($10\ {\rm days} < P < 1000\ {\rm days})$. Periodicity analysis is performed only on $g$- and $r$-band light curves; $i$-band photometry is provided in the Forced Photometry Catalog but is not searched for periods.

    \item[5.] {\bf Variability Metric Catalog}: Contains variability metrics run on all forced photometry difference fluxes.

    \end{itemize}

Table \ref{tab:accounting} summarizes the number of unique sources contained in the ZVAR data at various stages of analysis.

\begin{deluxetable*}{l l r r r r}
\tabletypesize{\footnotesize}
\tablewidth{0pt}
\tablecaption{Quantities reported for ZVAR DR1, with the definition of each and
the corresponding count. Counts are from an audit of the delivered files over the
635 primary-grid fields for which all matchfiles were readable. A \emph{usable
epoch} is one with a non-NaN \texttt{flux} and \texttt{flux\_err}. Dashes mark
quantities that are not resolved by band.\label{tab:accounting}}
\tablehead{
  \colhead{Quantity} & \colhead{Definition} &
  \colhead{$g$} & \colhead{$r$} & \colhead{$i$} & \colhead{Total}
}
\startdata
Unique sources &
  \parbox[t]{1.95in}{\raggedright
    Distinct PS1 IDs (\texttt{ps1\_id})\strut} &
  \nodata & \nodata & \nodata & 1{,}288{,}036{,}496 \\[3pt]
Source Catalog rows &
  \parbox[t]{1.95in}{\raggedright
    Distinct source--quadrant associations, one per
    (\texttt{ps1\_id}, \texttt{field}, \texttt{ccd},
    \texttt{quad})\strut} &
  \nodata & \nodata & \nodata & 1{,}466{,}645{,}547 \\[3pt]
Light curves &
  \parbox[t]{1.95in}{\raggedright
    Source--quadrant--band combinations with at least
    one usable epoch\strut} &
  1{,}398{,}706{,}231 & 1{,}397{,}115{,}051 &
  662{,}730{,}310 & 3{,}458{,}551{,}592 \\
\enddata
\end{deluxetable*}

\subsection{Source Catalog}
The Source Catalog provides basic information on all sources for which forced photometry has been performed.
ZTF observations are taken on two fixed grids of fields to simplify data processing, i.e., there is no explicit image dithering. The primary and secondary field grids are offset from each other by about $5^\circ$ diagonally to fill in CCD gaps \citep{Bellm_ZTF_overview}. The grids are aligned to cover the Galactic Plane with the fewest pointings but there is increasing overlap between neighboring fields with increasing Declination. A single position on the sky may therefore be covered by more than one ZTF field; however, the resolution for independent ZTF data reduction is one of the 64 readout channels (16 CCDs, each divided into 4 readout quadrants) that make up an observation. The Source Catalog reports each individual  ZTF association (field/CCD/quadrant) to a PS1 source within the ZTF primary grid. The file structure for the ZVAR DR1 Source Catalog is given in Appendix \ref{app:source-catalog}. 

\begin{figure}
    \centering
    \includegraphics[width=0.95\linewidth]{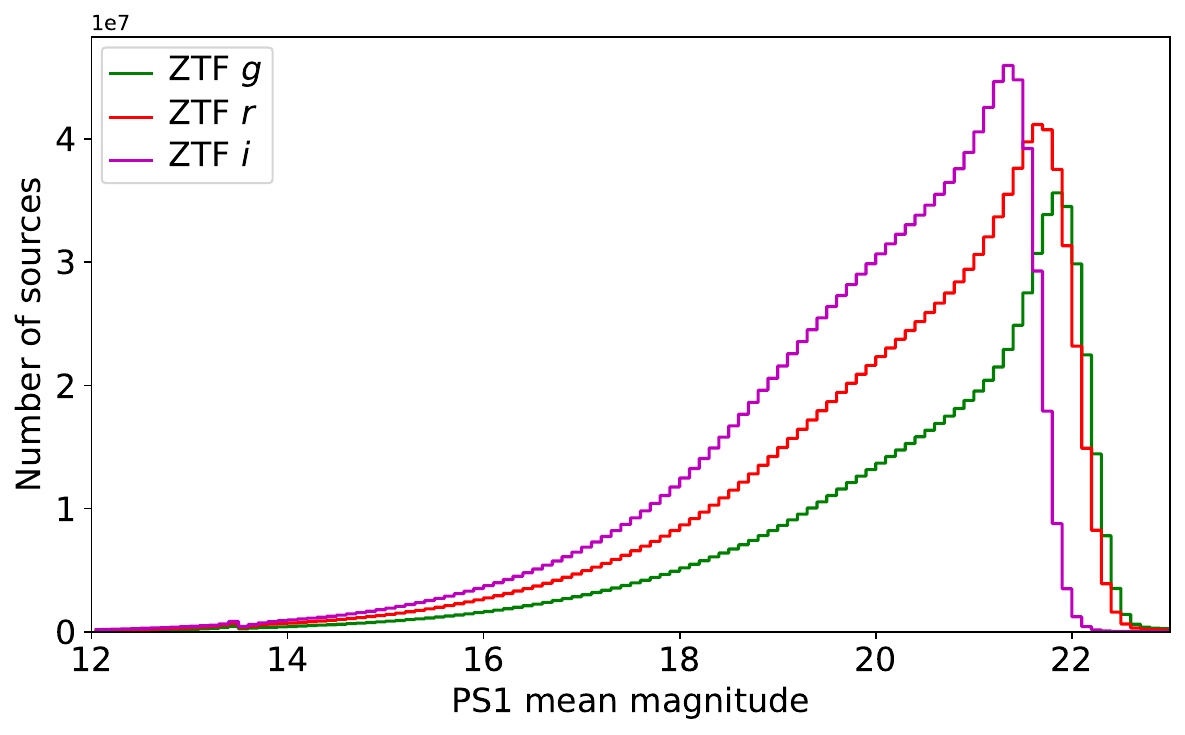}
    \caption{The respective ($g,r,i$) mean PS1 magnitude distributions for all objects in the Source Catalog.}
    \label{fig:magdist}
\end{figure}

\begin{figure}
    \centering
    \includegraphics[width=0.95\linewidth]{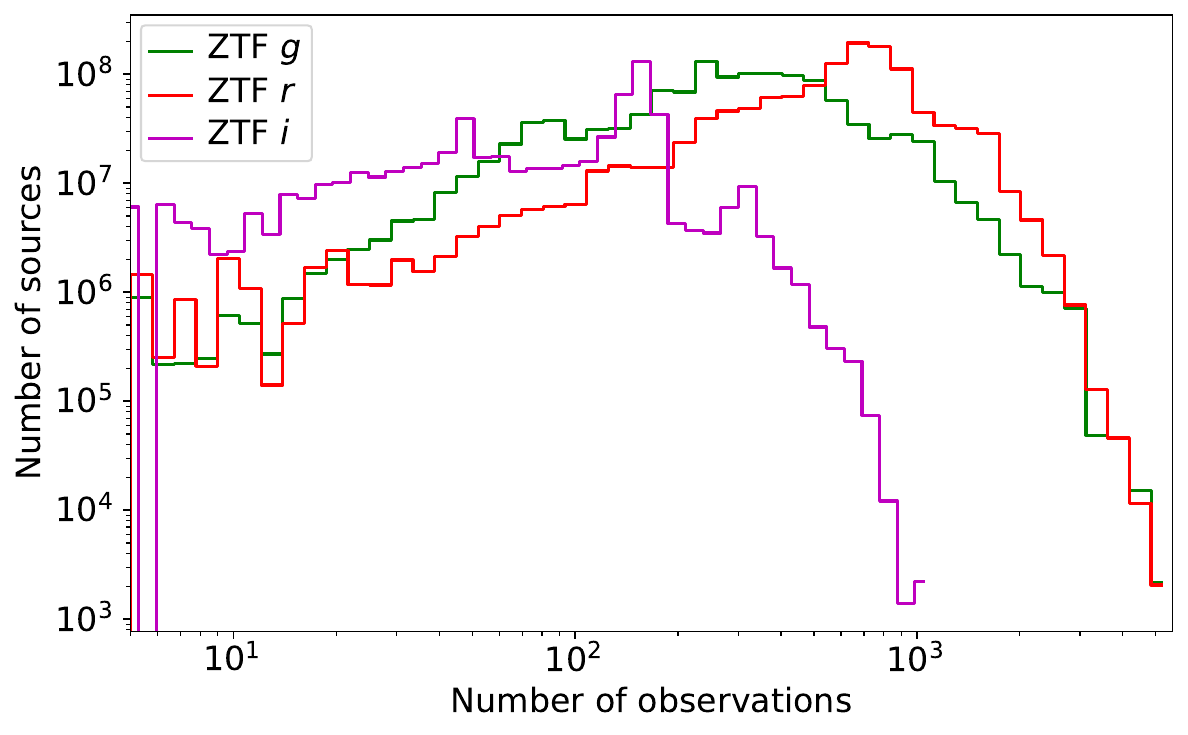}
    \caption{The respective ($g,r,i$) distributions for the total number of detections per source for all objects in the Source Catalog. Detections of a source in more than one ZTF quadrant have been summed so sources can have more observations than the number of times a specific field has been visited.}
    \label{fig:Ndist}
\end{figure}

\subsection{Forced Photometry Catalog}
Forced photometry is done on ZTF image difference data and involves inverse-variance-weighted PSF flux extraction for every object with a PS1 ID within the ZTF footprint on the sky.  
The photometry is given for \mbox{ZTF-$g$}, \mbox{ZTF-$r$}, and \mbox{ZTF-$i$} bands. \footnote{See
\url{https://github.com/ZwickyTransientFacility/ztf_information} for definition of
the filter transmission functions.}
ZVAR DR1 contains photometry exposures from March 19, 2018 to July 31, 2023. 
The Forced Photometry Catalog contains 1.29 billion sources (unique PS1 ids).
The median number of observations among sources with at least one observation in that band is: $n_r=636$, $n_g=286$, and $n_i=108$ (see Fig.~\ref{fig:Ndist}).

\begin{figure*}
    \centering
    \includegraphics[width=0.95\linewidth]{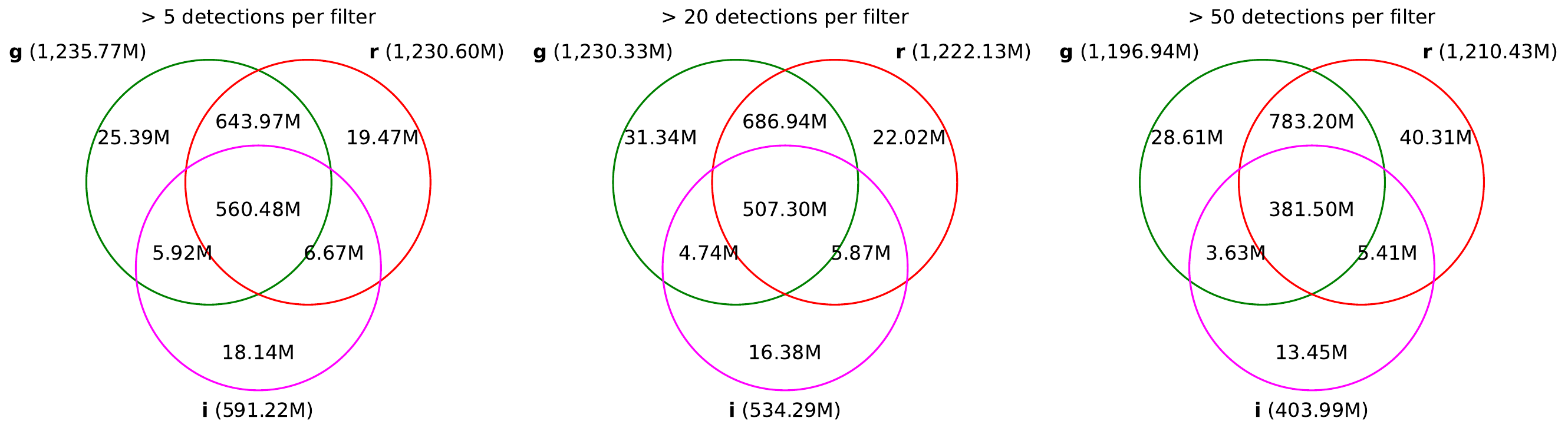}
    \caption{The overlap of sources with more than 5, 20, and 50 detections in the ZTF $g$, $r$, and $i$ filters.}
    \label{fig:venncoverage}
\end{figure*}

\begin{figure*}
    \centering
    \includegraphics[width=0.49\linewidth]{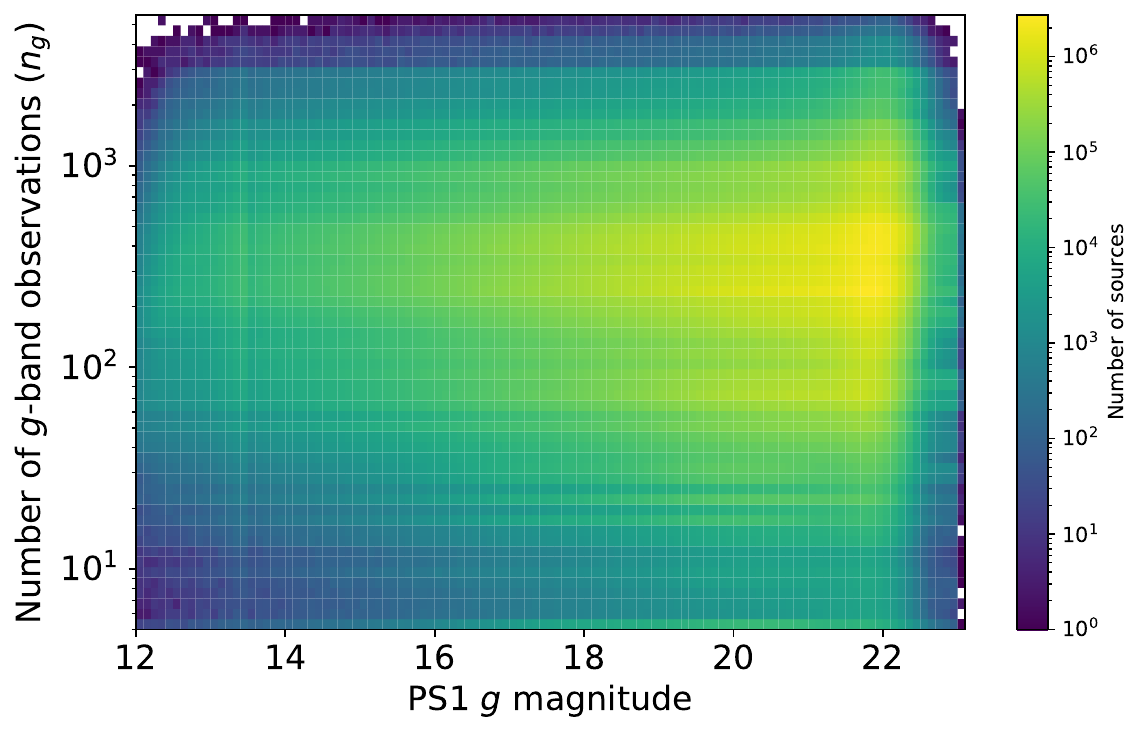}
    \includegraphics[width=0.49\linewidth]{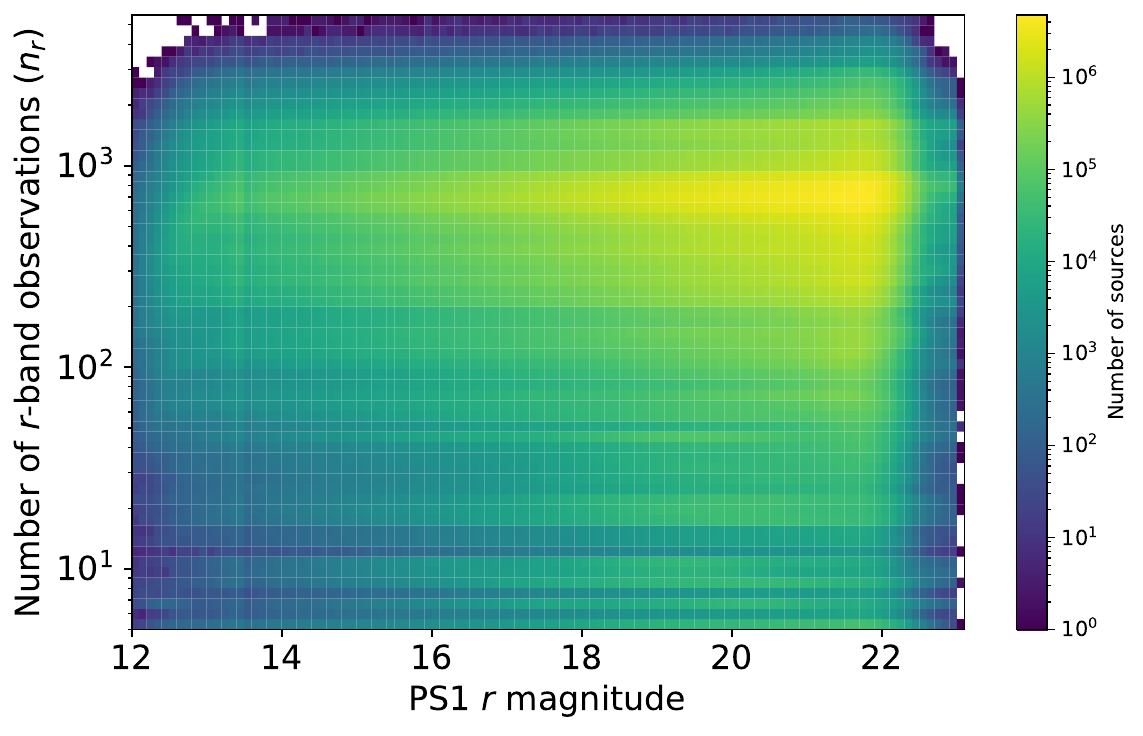}
    \caption{The distribution of the total number of observations of a source as a function of (left) $g$- and (right) $r-$ magnitude  for all objects in the Source Catalog. Detections of a source in more than one ZTF quadrant have been summed.}
    \label{fig:nmagdist}
\end{figure*}

ZTF observes the sky above a minimum declination of $\delta > -30^\circ$.
    The number of exposures in each field for ZVAR DR1 is shown in figure \ref{fig:coverage}.
    This coverage is very heterogeneous due to several overlapping survey strategies and a priority-based observing scheduler \citep{Bellm19_zwicky_transient_schedule}.
        ZVAR DR1 covers the ZTF primary grid field \citep{Bellm_ZTF_overview}\footnote{ZTF also has a secondary field grid which covers the gaps between primary fields but this is not included in ZVAR DR1.
        See \url{https://github.com/ZwickyTransientFacility/ztf_information} for details on the ZTF field grid.}
        (See Section~\ref{sec:discussion} for further details.)

A source can appear in more than one forced photometry file because of
overlaps in the ZTF primary grid fields. 
These are treated as independent light curves in the periodicity
analysis  and in construction of the variability metrics.  
Documentation on the pipeline for producing ZVAR image-difference forced photometry
can be found at \url{https://github.com/zvar-astro/ZVAR-Forced-Phot}.

\subsection{Periodicity Catalogs}
Periodograms are computed for the $g$- and $r$-band forced photometry light curves of objects in the ZTF primary grid using the Fast Periodicity Weighting (FPW) algorithm
\citep{finkbeiner2025_FPW}. 
Candidate periodicities are identified for objects having at least 50 epochs in either the $g$- or $r$-band. 
Because ZTF obtains substantially fewer $i$-band than $g$- and $r$-band epochs ($i$-band usage has been programmatically limited), periodograms are not computed for $i$. Note that epochs from ZTF deep-drilling observations are not used in this periodicity analysis, i.e., observations of the same field with less than 5 minutes between successive visits. The FPW periodogram is a phase binning statistic defined for a trial frequency, $f$, as:

\begin{large}
\begin{equation}
    \chi^2_{\rm FPW}(f) \equiv \sum_{m=1}^{\Nbin} \left[ 
    \frac{\left( \sum_{j=1}^{N_m}  \frac{f_{m,j}}{{\sigma_f^2}_{m,j}} \right)^2}
   {\sum_{j= 1}^{N_m} \frac{1}{{\sigma_f^2}_{m,j}}} \right] ~,
   \label{eq:phasesum}
\end{equation}
\end{large}

\noindent
where $\Nbin$ is the number of phase bins, $N_m$ is the number of observations in the $m$th phase bin, $f_{m,j}$ is the mean-subtracted difference flux of the $j$th photometric observation in the $m$th phase bin, and ${\sigma_f^2}_{m,j}$ is the squared observation error of $f_{m,j}$. 

ZVAR DR1 uses three configurations of phase binning: $N_{bin} =$ 20, 10, and 5 bins.
 The binning configurations determine how sensitive the periodogram is to different waveforms with 5 bins being most sensitive to sinusoids and 20 bins being most sensitive to sharp features such as eclipses (see \citet{finkbeiner2025_FPW} for a detailed description). Trial frequencies are sampled on a linear grid with spacing $\Delta f= 1/(\alpha T)$, where $T$ is the time span of the quadrant light curve file with an oversampling of $\alpha$. 

 Periodicity catalogs were run separately for the SPC and LPC due to different data processing required for the two searches, in particular, the degree of oversampling of the frequency grid (see Table~\ref{tab:periodpars}) and the removal of a long time baseline for the SPC. For each source, the respective periodicity catalog contains a number of the most significant periods per binning configuration. 

\begin{table*}
\centering
\small
\caption{Parameters for the periodicity search in ZVAR DR1}
\label{tab:periodpars}
\begin{tabular}{lcc}
\hline\hline
Parameter  & Short Period Catalog (SPC) & Long Period Catalog (LPC)\\
\hline
Period range & 5m -- 10d & 10d -- 1000d \\
Oversampling ($\alpha$) & 3 & 300 \\
No. of significant periods per binning & 50 & 32 \\
Median significant periods per source: & & \\
\hspace{0.2in}$N_{bin} = 20$ & 3 & 1 \\
\hspace{0.2in}$N_{bin} = 10$ & 15 & 1 \\
\hspace{0.2in}$N_{bin} = 5$ & 50 & 1 \\
Minimum significance threshold $(z)$ & 7 & 10 \\
\hline
\end{tabular}
\end{table*}

 The {\it significance} of a periodogram peak is defined to be the number of standard deviations a specific periodogram point is above the mean,
i.e., 
the significance, $z(f)$, of a peak is defined to be

\begin{equation} \label{eq:significance}
    z(f) = \frac{\chi^2(f) - \mu_{\chi^2}}{\sigma_{\chi^2}}
\end{equation}

\noindent where $\chi^2(f)$ is the periodogram value at test frequency $f$ and $\mu_{\chi^2}$ and $\sigma_{\chi^2}$ are the mean and standard deviation of the periodogram, respectively. A peak is only considered a candidate if it falls above a minimum significance threshold. 
In searches for short period objects with $P_{\rm max} \sim 10$ days, the use of a simple mean and standard deviation is adequate as the majority of test frequencies are uncorrelated. In the long period analysis, test frequencies become significantly correlated so we calculate a sigma-clipped mean and standard deviation using a $3 \sigma$ cutoff.

The distribution of significance values, $z(f)$, varies with binning and the number of observations per source 
and must be calibrated to a probability that a peak could arise by
chance from a random phasing of a light curve.
This is done by an empirical fit of the significance distribution to a
generalized noise distribution.
The details of the conversion from significance, $z(f)$, to a false alarm probability (FAP) with a white noise null hypothesis are given in
\url{https://github.com/zvar-astro/ZVAR-Period}.
The ZVAR DR1 API provides a conversion from significance to
a FAP and a further conversion to the equivalent number of Gaussian standard deviations, $\sigma_{FAP}$, a periodogram peak is above the noise distribution.

 We note, however, that a peak with high significance, or equivalently
 low false alarm probability, does not necessarily indicate
a true periodicity in the source. Regularities in the times at which photometry is sampled
(e.g., sidereal day, lunar month, year), false positive peaks
may appear in the periodogram at relatively high significance
\citep[see, e.g.][]{VanderPlas_LS}.  
In ZVAR DR1, certain period ranges are therefore excised
from consideration in the FPW period analysis: for example, 
the sidereal day frequency shows a strong peak and integer multiples and divisors of it 
can also be strong. A list of excised period ranges is given in Appendix \ref{app:excise}. We note, 
however, that these excisions are not exhaustive and higher order harmonics may still be found in the ZVAR SPC and LPC.


We also apply an additional mitigation of spurious peaks by fitting a negative binomial to the distribution of peaks in each field-CCD-quadrant-filter set and remove frequencies that share more peaks between independent sources than the $3\sigma$ quantile threshold. Frequencies above a median threshold of 6 shared peaks are blacklisted.

False positives are a particularly common problem with periodicity searches as the peak in the periodogram
may not represent the true period of the source (see, e.g., \cite{VanderPlas_LS}, Figure 24): 
for example, a periodogram peak can be at a beat frequency of the true source frequency and
the frequency of a sampling regularity.  A source with primary and secondary eclipses may also have its 
most significant period at half the true period.

{\it We advise therefore that the periodicities in the ZVAR SPC and LPC should be treated as candidate periods
rather than as definitive true periods of the source.}
Additional confirmation of periods with high-speed photometry, detailed phase folding, or
other techniques is recommended.
We will discuss the techniques and the underlying rationale of the ZVAR
periodicity analysis in more detail in a future paper 
(Whitebook et al., in prep.).


\subsection{Variability Metric Catalog}\label{sec:metrics}

ZVAR DR1 includes a number of statistical variability metrics based on the photometric data. These statistics can provide the basis for automated classification and characterization of different variability types via machine learning techniques, e.g., SCoPe \citep{Healy_SCOPE}. However, the individual metrics can also provide an assessment of underlying variability when compared to their statistical distributions.

\begin{table}[h]
\centering
\small
\caption{ZVAR variability metrics included in DR1.}
\label{tab:metrics}
\begin{tabular}{ll}
\hline\hline
Metric & Description\\
\hline
$\sigma$ & \texttt{standard deviation}\\
$p2p$ & \texttt{point-to-point scatter}\\
$\rm MSSD$ & \texttt{mean square successive difference}\\
$S$ & \texttt{skew (third standarized moment)}\\
$S_p$ & \texttt{Pearson's skew}\\
$\kappa$ & \texttt{kurtosis (fourth standarized moment)}\\
$\sigma^{2}_{NXS}$ & \texttt{normalized excess variance}\\
$\eta$ & \texttt{von Neumann ratio}\\
$IQR$ & \texttt{interquartile range}\\
\hline
\end{tabular}
\end{table}

The ZVAR variability metrics provided in this release are given in Table \ref{tab:metrics}. 
They are calculated based on \emph{difference fluxes}, rather than magnitudes, and so special care is required in interpretation or comparison with other data sets. Additionally, ZVAR metrics are only calculated for ZVAR light curves with at least 20 data points after selecting difference flux values within a 10 -- 90th percentile range. Thus, many of the faintest objects do not have corresponding values for these metrics. Here we provide details for the less frequently used parameters.

The average point-to-point scatter is the variation between pairs of points taken in order:
\begin{equation}
\label{eq:p2p}
p2p=\sqrt{\frac{\sum^{n-1}_{i=1}{(f_{i+1}-f_{i})^2}}{n-1}}
\end{equation}\\
for $n$ difference flux measurements $f_i$.

The normalized excess variance is given by:
\begin{equation}
\label{eq:nxs}
\sigma^2_{NXS}=\frac{1}{n\bar{f}^2}\sum^n_{i=1}[(f_i-\bar{f})^2-\sigma_i^2]
\end{equation}\\
where $\sigma_i$ are the errors of $n$ individual difference flux measurements.
This statistic is invariant under multiplicative flux calibration and can serve as a measure of variability characterization. It is often used for objects exhibiting red noise or random walk processes, such as AGN.
It is also useful for understanding the accuracy of the uncertainties, since negative values suggest that the photometric errors are larger than the scatter, while positive values suggest that they vary more than expected if the errors are correct. It should be noted that the value can become numerically unstable when the 
the mean difference flux approaches zero.

The von Neumann ratio \citep{vonNeumann_1941} is equivalent to the MSSD / variance, where the MSSD is the point-to-point scatter squared ($p2p^2$), that is:

\begin{equation}
\label{eq:vn}
\eta = \frac{\sum^{n-1}_{i=1}(f_{i+1}-f_i)^2/(n-1)}{\sigma^2}
\end{equation}




\section{Query Tools and Data Availability} \label{sec:data_avail}
ZVAR DR1 data can be accessed directly from an API at \url{https://api.zvar.caltech.edu/imdiff} \footnote{A full list of endpoints with examples can be viewed online at \url{https://api.zvar.caltech.edu/imdiff/docs}} or viewed at \url{https://zvar.caltech.edu}. Login credentials are required and instructions on how to obtain these can be found at \url{https://zvar.caltech.edu/help}. The API has separate methods (or "endpoints") for retrieving photometry, periods, variability metrics and cross-matching. 

To retrieve the image difference forced photometry directly, the \texttt{/photometry} method takes in a list of \psid s and optional restrictions to return a subset of \textit{g, r, i} light curves, or to return the photometry in flux or magnitudes. By default, $g$ and $r$ data is returned and the photometry is formatted as AB magnitudes calibrated to the PS1 DR1 system. The magnitudes are determined by converting the PS1 source magnitude to a reference flux using the appropriate zeropoint and color term and then adding this to the measured image difference fluxes. These are then converted back to an AB magnitude used the same zeropoint and color term. The photometry method may return more than one light curve per band for some objects if there is spatial overlap between two different ZTF fields. An example query is shown in listing \ref{lst:photometry_query}.

\begin{lstlisting}[language=Python, basicstyle=\small\ttfamily,
    keywordstyle=\ttfamily,
    caption={Example \texttt{Python} query to the ZVAR IMDIFF API \texttt{/objects}
    endpoint, retrieving all objects with periods between 5 minutes
    and 30 minutes and a significance $>20$.},
    label={lst:objects_query}]
import requests
payload = {"min_period": 5 * 60,
           "max_period": 30 * 60,
           "min_significance": 20.0}
response = requests.post(
    "https://api.zvar.caltech.edu/imdiff/objects",
    auth=(username, password),
    json=payload)
data = response.json()
psids = [obj["psid"] for obj in data]
periods = [obj.get("periods", []) for obj in data]
\end{lstlisting}

\noindent
where \texttt{min\_period} and \texttt{max\_period} are in seconds,
\texttt{min\_significance} is the minimum significance defined
in Equation \ref{eq:significance},
and the returned \texttt{periods} are in days.

Objects can be found with a cone search method, \texttt{/cone\_search}, which will return the \psid\ of each object within a coordinate search with optional user constraints, such as a range of periods or a minimum significance that at least one period for the object satisfies. A single object can be searched for with its \psid, \texttt{/object/\{psid\}}. A user may also search for every object satisfying some period specific criteria with the general \texttt{/objects} method, such as every object with a period between $5 - 30$ minutes with significance greater than $20$, which will retrieve many white dwarf pulsators and ultracompact binaries. We show an example call to the \texttt{objects} method in listing \ref{lst:objects_query}.

The API can be used to query ZVAR objects with possible external catalog ids using the \texttt{/external\_xmatch} method. The supported catalogs are: \textit{AllWISE, eROSITA, Gaia, Galex, Pan-STARRS, 2MASS}, and \textit{VSX}. 
ZVAR objects within a catalog can be returned by specifying which catalog to use as a base, a list of ids for that catalog, and a search radius in arcseconds. This cross match can also be inverted: given a list of \psid s from ZVAR, properties from matching objects in external catalogs can be retrieved with the \texttt{/xmatch} method. The list of parameters returned from each possible catalog is shown in Table \ref{table:xmatch}.

\begin{deluxetable}{lll}
\tablecaption{Data products returned by cross-match with each catalog.
All catalogs return a source identifier and the coordinates of each match.
All magnitudes are returned with their uncertainties.\label{table:xmatch}}
\tablewidth{0pt}
\tablehead{
  \colhead{Query Keyword} & \colhead{Returned quantities} & \colhead{System}
}
\startdata
\texttt{AllWise}\tablenotemark{a} &
  $W1, W2, W3, W4$ & Vega \\
\texttt{TwoMass}\tablenotemark{b} &
  $J, H, K_s$ & Vega \\
\texttt{PanSTARRs}\tablenotemark{c} &
  $g, r, i, z, y$ & AB \\
\texttt{Galex}\tablenotemark{d} &
  FUV, NUV & AB \\
\texttt{Gaia}\tablenotemark{e} &
  \parbox[t]{3.6cm}{\raggedright
    $G, G_{BP}, G_{RP}$;
    $\varpi, \mu_\alpha, \mu_\delta$\strut} &
  Gaia \\[3pt]
\texttt{ERosita}\tablenotemark{f} &
  0.2--2.3\,keV flux & cgs \\
\texttt{VSX}\tablenotemark{g} &
  \parbox[t]{3.6cm}{\raggedright
    Variability type, period, mag range\strut} &
  mixed \\
\enddata
\tablecomments{%
$^{\rm a}$AllWISE \citep{Wright_WISE,Mainzer_NEOWISE};
$^{\rm b}$2MASS Point Source Catalog \citep{Skrutskie_2MASS};
$^{\rm c}$Pan-STARRS1 DR1 \citep{Chambers_PS1,Flewelling_PS1};
$^{\rm d}$GALEX \citep{Martin_GALEX};
$^{\rm e}$Gaia EDR3 \citep{GaiaMission,GaiaEDR3}, astrometry from
\citet{Lindegren_EDR3_astrometry};
$^{\rm f}$eROSITA-DE DR1 (eRASS1) \citep{Merloni_eRASS1}, flux in
erg\,s$^{-1}$\,cm$^{-2}$;
$^{\rm g}$AAVSO International Variable Star Index \citep{Watson_VSX},
version of 2025 October 19; magnitudes are in the band listed for each entry.}
\end{deluxetable}


\begin{lstlisting}[language=Python, basicstyle=\small\ttfamily,
    keywordstyle=\ttfamily,
    caption={Example \texttt{Python} query to the ZVAR IMDIFF API
    \texttt{/photometry} endpoint, retrieving the \textit{g}- and
    \textit{r}-band light curves of two objects in magnitudes.},
    label={lst:photometry_query}]
import requests
payload = {"psids": [psid_1, psid_2],
           "bands": ["g", "r"],
           "format": "mag"}
response = requests.post(
    "https://api.zvar.caltech.edu/imdiff/"
    "photometry",
    auth=(username, password),
    json=payload)
lightcurves = response.json()
lc = lightcurves[str(psid_1)]
g = [p for p in lc if p["band"] == "g"]
times = [p["time"] for p in g]
mags = [p["mag"] for p in g]
\end{lstlisting}

The key for the returned JSON object is \psid. Each value is a single array of epochs, containing every requested band and sorted by ascending observing time. Objects with no photometry on the requested bands are omitted. Each epoch is a flat record containing the fields described in Table~\ref{table:photometry}. Values that are undefined are returned as \textit{null}.

\par\vspace{10pt}
\noindent\begin{minipage}{\linewidth}
\refstepcounter{table}\label{table:photometry}
{\small \textbf{Table \thetable.} Fields of each epoch returned by the \texttt{/photometry} method.\par}
\vspace{6pt}
\centering
\footnotesize
\begin{tabular}{@{}ll@{}}
\hline\hline
Key & Description \\
\hline
\texttt{time}                               & Barycentric Julian Date (d) \\
\texttt{band}                               & ZTF filter (\textit{g}, \textit{r}, \textit{i}) \\
\texttt{mag}, \texttt{magerr}               & PS1-calibrated AB mag and $1\sigma$ error \\
\texttt{limmag}                             & $5\sigma$ limiting AB mag (non-detections) \\
\texttt{flux}, \texttt{fluxerr}             & Difference-image flux and $1\sigma$ error (DN) \\
\texttt{snr}                                & Signal-to-noise ratio \\
\texttt{field}, \texttt{ccd}, \texttt{quad} & ZTF field, CCD and quadrant \\
\texttt{infobits}                           & ZTF exposure quality bitmask \\
\hline
\end{tabular}
\end{minipage}
\par\vspace{10pt}

\section{Illustrative Science Result} \label{sec:results}
 In the ZVAR DR1 SPC, we have identified 7,829,930\, periodic light curves with Gaussian equivalent false alarm probabilities (\FAPsig) above 3$\sigma$, 5,997,936\, with \FAPsig $> 5 \sigma$, and 2,714,273\, with \FAPsig $> 7 \sigma$. Correspondingly in the LPC, we find 1,266,098\, with \FAPsig $> 3 \sigma$, 170,856\, with \FAPsig $> 5 \sigma$ and 61,294\, with \FAPsig $> 7 \sigma$. 

\begin{figure*}[t!]
    \centering
    \includegraphics[width=\linewidth]{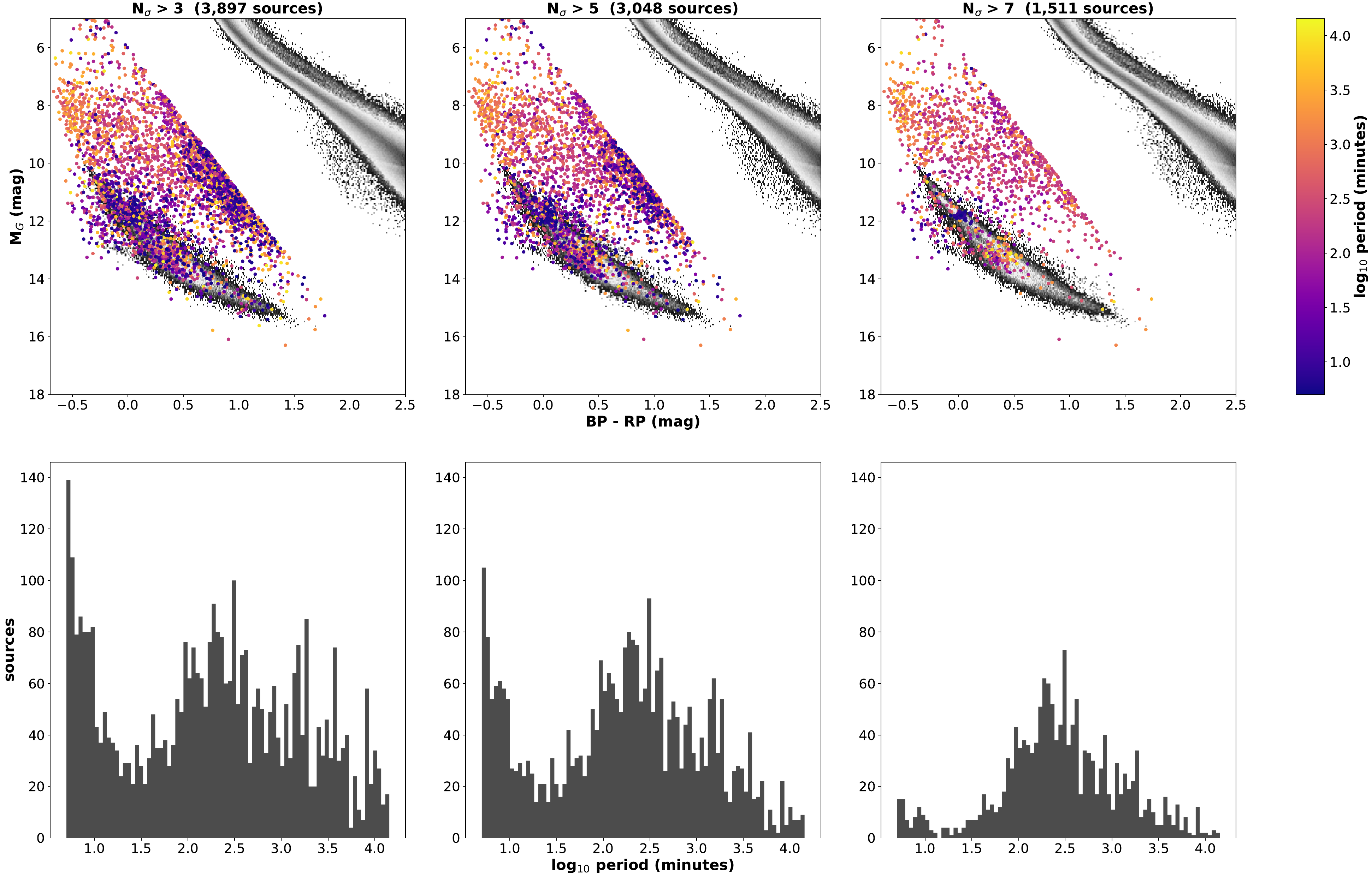}
    \caption{Top: HR diagrams with a scatter of periodic variable white dwarfs from the Gaia EDR3 sample \citep{Gentile_Fusillo_2021} found in ZVAR with \FAPsig $>3\sigma$, $>5\sigma$, and $>7\sigma$ colored by their most significant period and overlaid on an HR diagram of the Gaia 200 pc sample in gray. Bottom: The distribution of the most significant periods per object. We note a pile-up of sources at the lowest periods in the low \FAPsig limit, which is common for variable, but non-periodic stars as high frequencies provide the most chances for false alarms. A decrease of objects between 2.15 and 3.18 hours corresponding to the cataclysmic variable period gap is also apparent.}
    \label{fig:WD_distribution}
\end{figure*}

To demonstrate the potential of these data, we crossmatched the Gaia EDR3 white dwarf (WD) catalog \citep{Gentile_Fusillo_2021} with ZVAR using a 2'' crossmatch radius. 566,987 have a match in ZVAR which represents 88.7\% of the Gaia EDR3 WD sample within the ZTF footprint; the majority of the missing $\sim$11\% fall within tiling gaps between ZTF CCDs. From the SPC, 
those with at least one significant period at $>3, 5, 7 \sigma$ are: $N(\sigma_{FAP} > 3\sigma) = 3897 (0.69\%)$, $N(\sigma_{FAP} > 5\sigma) = 3048 (0.54\%)$, and $N(\sigma_{FAP} > 7\sigma) = 1511 (0.27\%)$. The distribution of periods in each cut and a color-magnitude distribution of the periodic WDs found in ZVAR is shown in Figure \ref{fig:WD_distribution}.

Comparing with known variable WDs from the Variable Star Index \citep[VSX;][]{Watson_2006} and SIMBAD databases, $811$ ($20.4\%$) of our $3 \sigma$ candidates are previously known variables with $452$ objects in both VSX and SIMBAD, $359$ only in VSX, and $30$ only in SIMBAD. As the $N_\sigma$ cuts become stricter, the fraction of ZVAR periodic variable candidates that are previously known increases, as expected: $N(\sigma_{FAP} > 5 \sigma)$ $752$  ($25\%$) and $N(\sigma_{FAP} > 7 \sigma)$ $556$ ($37\%$) are previously known, respectively. Conversely, of the 566,987 Gaia EDR3 WDs within the ZVAR footprint, 2878 are found in VSX. Recovery of WDs in VSX by ZVAR is a strong function of the periodicity class: many objects in VSX are stochastic variables with no long-term coherent periodicity e.g., dwarf novae, and others are multiperiodic low-amplitude pulsators, e.g., ZZ Ceti and V1093 Her. These specific classes are difficult for periodograms to detect. 

We report recovery of VSX known variable WDs within the ZVAR footprint both cumulatively and split into classes: coherent, stochastic, and multimode. Coherent periodic sources are those classes which have a single periodic mode. These include all eclipsing sources, reflection and ellipsoidal effects, synchronous rotators such as polars, and single-mode radial pulsators. Stochastic variables are eruptive variables, some of which possess an underlying coherent periodic signal. Multimode variables possess multiple periodic waveforms. Of the 2870 VSX variable WDs in the ZVAR footprint, ZVAR recovers $811$ ($28\%$). ZVAR recovers $426$ of the $686$ coherent systems in VSX ($62\%$), $176$ of $796$ stochastic variables ($22\%$), $37$ of $522$ multimode variables ($7\%$), and $172$ of $866$ unclassified variables ($20\%$). The majority of missed sources are detected as periodic in ZVAR, but fall below the $3 \sigma$ cut.

The majority of VSX sources are from surveys that are magnitude-limited brighter than ZVAR. The median Gaia G band magnitude of VSX matching WDs is $19.7$, while the median Gaia G band magnitude of ZVAR periodic variables from the same sample is $20.2$. We note that this comparison is not between two fully independent samples. Many of the variables cataloged in VSX were themselves found in ZTF prior to the ZVAR survey.

\section{Discussion} \label{sec:discussion}

Variable source studies have greatly benefited in recent years from the release of a number of large catalogs of time series by sky surveys, and systematic searches through these for particular classes of variable source,. However, these are limited to source detections at a nominal survey detection limit, $5 \sigma$, say, and so focus on the brighter, closer populations. Sparsely sampled light curves of fainter sources have carried insufficient information for anything but the most cursory of characterization except, perhaps, for the largest amplitude variable objects.
ZVAR DR1 represents the first large-scale forced photometry catalog generally available to the community and a set of light curves with the same local sampling irrespective of source brightness. As both detection and characterization sensitivity of a variable signal will typically improve approximately as the square root of the number of observations, this enables broader studies of fainter populations. 

ZTF is itself, however, an inhomogeneous data collection since it is the combination of public sky surveys and numerous boutique private surveys\footnote{ZTF has been jointly funded by the NSF and the ZTF Collaboration and usage is allocated accordingly.}. The celestial sky was mapped onto two fixed field grids \citep{Bellm_ZTF_overview}\footnote{\url{https://github.com/ZwickyTransientFacility/ztf_information}} and between 2018 and 2020, every field that was visible to ZTF at any given time was surveyed every 3 nights in $g$ and $r$ within the same night at least 30 minutes apart. The visible Galactic Plane within Dec $> -30^\circ$ and $|b| \le 7^\circ$ was visited twice each night in $g$ and $r$, again at least 30 minutes apart. Priority in both cases was given to the primary field grid pointings and ZVAR DR1 only contains data from those fields. From 2021 onwards, the northern sky survey cadence changed to every two nights. This results in a large variance in the number of exposures observed per field, e.g., low declination fields may have around 300 exposures whereas the most sampled field has over 3500, but the median sampling is 286 observations in $g$ and 636 in $r$.

The ZTF field-of-view is 47 deg$^2$ which translates to a physical focal plane size of approximately 400mm by 400mm. The shutter operates at a finite speed ($\sim$0.3 - 0.4s to open or close) which means that exposure times at the center and the edges of a frame are not exactly equivalent. The observation time recorded by ZTF is the time at which the signal to open is sent to the camera. The effects of this on periodicity analysis introduce at most a 0.15\% systematic phase uncertainty at the shortest periods tested.

Aside from these two characteristics of basic ZTF data,
the construction of the ZVAR data set also imposed a number of data systematics from the particular set of design decisions, including:\\
{\it Choice of background source catalog:} Forced photometry requires a set of positions at which measurements are made. For completeness of known sources (transient sources can have data at a position where no prior known source is known, e.g., a flaring M-star which is below a survey limit in quiescence), a source catalog from a deeper survey should be used. The Pan-STARRS DR1 (PS1) catalog spatially overlaps with ZTF and has limiting magnitudes of $g=23.3$, $r=23.2$, and $i = 23.1$ respectively, which are least 0.5 magnitudes dimmer than the limiting magnitude of ZTF. It also has sharper resolution: 0.257"/pixel with a median FHWM in $r$ of 1.19" \citep{Flewelling_PANSTARRS} compared to 1"/pixel and a median FWHM of 2" in $r$ for ZTF. Both PS1 and ZTF are calibrated to Gaia DR1 with astrometric accuracies of 20 and 65 milliarcseconds (30 milliarcseconds for bright stars), respectively.

It is possible that two sources resolved in PS1 will appear as a blended source in ZTF/ZVAR. In the case where one of the two is also variable, the measured difference flux will be affected by the increased background noise due to the subtracted non-variable source. Unfortunately, the difference image photometry in ZVAR is performed one object at a time rather than jointly across all sources in the field. This means that objects very close to a variable source tend to be affected to some extent by the varying background within the PSF-fit aperture. For users interested in the photometry of close $(<1")$ pairs of objects, the results should therefore be inspected and joint fitting considered.

The use of only ZTF primary grid data may also contribute to potential completeness issues. The ZTF AGN catalog \citep{Nackoneczny_QZO} contains 4359418 sources with unique PS1 ids which are distributed uniformly across the ZTF footprint which makes it a good test data set. The ZVAR Source Catalog contains 4141071 of these and 123122 of the missing sources do not lie within the footprint of the primary grid. The rest of the missing sources are preferentially located in the Galactic Plane, have between 20 and 30 detections, and/or are fainter than $g \sim 20$. We estimate the completeness
 of the ZVAR Source Catalog as $\sim$98\%.

\noindent
{\it Choice of forced photometry algorithm:} For a given difference image, the ZTF Forced Photometry Service \citep[FPS,][]{Masci2023} employs a local position dependent PSF model precalculated for that specific exposure frame. The amplitude of the PSF is then optimized at the target coordinates against local pixel values using linear least-squares minimization. This calculation is then repeated across the hundreds of difference images that cover the target position. ZVAR uses the same difference images but fits a PSF model at the target position through bilinear interpolation. The measured flux is then the inverse-variance weighted flux rescaled to a fiducial zero point for the image. No further postprocessing is done. The output from FPS and ZVAR has been found to be broadly comparable (Vanderbosch, 2024, priv. comm.) and a more detailed comparison is beyond the scope of this paper.

\noindent
{\it Choice of periodicity algorithm:} It has been demonstrated that there is no ideal period finding algorithm and that the relative accuracy of different algorithms depends on data properties, such as signal-to-noise ratios, sampling, and cadence, as well as object class \citep{Graham2013}. Ensemble techniques do not fare any better. The fixed field grid that ZTF uses for its observations means that all sources within a specific field share individual timestamps, which are all central frame mid-exposure BJDs. The BJDs are accurate to $\sim$3 seconds with systematic variations depending on source location and season. The shared timestamps defines a natural batch processing for all light curves within the same field as phase folding operations at trial frequencies only need to be calculated once per field. FPW is an algorithm which can utilize this in a GPU deployment for efficient bulk period finding. Further, FPW is a waveform agnostic algorithm capable of detecting different classes of variables simultaneously. A degree of post processing is required to interpret the results but a more detailed discussion of this is deferred to Whitebook et al. (in prep.).

\noindent
{\it Choice of variability metrics:} Variability measures are a common constituent of feature sets for automated machine learning classifiers trained on and applied to large sky surveys. There is, however, realistically only a finite amount of (statistical) information that can be extracted from most time series appearing in such data sets and the sets of tens to hundreds of different features used will inevitably be highly correlated. Feature relevance often appears as a post classification analysis but comparative analyses of the most effective variability measures are rare, e.g., \cite{SokolovskyWelchI2017}. We have limited the choice of variability metrics that we calculate per source to the most informative from literature reports. Additional considerations were the computation time over 1.5 billion time series and the size of the final variability metrics catalog.

{\it Correction for systematic effects:} All ZTF data processing and published photometry is performed at the resolution of a readout channel (1/64th of the full field-of-view) with no corrections for systematic effects above that scale, e.g., spatial gradients across the full field-of-view. No corrections are applied either at the single quadrant level. This applies as well to all forced photometry and we have not corrected any data in DR1.

As has been mentioned, the second ZVAR data release (DR2) will feature forced photometry of the science images. Science image photometry is generally of better quality in areas of low crowding and is of particular utility in the search for variable high proper motion objects since proper motions are accounted for (they are not in the difference image forced photometry). Joint FP fits are also performed with the science images which will deal with some of the blending issues discussed above but is still not ideal in very crowded fields as there are often additional blended sources that are uncataloged. 
DR2 will also employ the ZTF Deep Reference Catalog as its seed which goes fainter than PS1 and so will report photometry for more sources.

\section{Software and third party data repository citations}
\label{sec:cite}
The basic user documentation available for ZVAR is summarized in Table~\ref{tab:document}.

\begin{table*}
\centering
\caption{Online material available for ZVAR DR1}
\label{tab:document}
\begin{tabular}{lll}
\hline\hline
\parbox[t]{1.1in}{\raggedright Subject\strut} &
  \parbox[t]{2.3in}{\raggedright URL\strut} &
  \parbox[t]{\dimexpr 7in-1.1in-2.3in-6\tabcolsep\relax}{\raggedright Description\strut} \\[3pt]
\hline
\parbox[t]{1.1in}{\raggedright ZVAR webapp\strut} &
  \parbox[t]{2.3in}{\raggedright \url{https://zvar.caltech.edu}\strut} &
  \parbox[t]{\dimexpr 7in-1.1in-2.3in-6\tabcolsep\relax}{\raggedright The web interface for accessing ZVAR DR1 data.  It contains general documentation as well as tools for source selection, periodicity selection, and various cross match information.\strut} \\[3pt]
\parbox[t]{1.1in}{\raggedright ZVAR API\strut} &
  \parbox[t]{2.3in}{\raggedright \url{https://api.zvar.caltech.edu/docs}\strut} &
  \parbox[t]{\dimexpr 7in-1.1in-2.3in-6\tabcolsep\relax}{\raggedright The programmable API that can be used for selection of ZVAR DR1 sources and cross match information.\strut} \\[3pt]
\parbox[t]{1.1in}{\raggedright ZVAR V1 build\strut} &
  \parbox[t]{2.3in}{\raggedright \url{https://zvar.caltech.edu/time-domain-analysis}\strut} &
  \parbox[t]{\dimexpr 7in-1.1in-2.3in-6\tabcolsep\relax}{\raggedright DR1 Details\strut} \\[3pt]
\parbox[t]{1.1in}{\raggedright ZVAR Forced Photometry\strut} &
  \parbox[t]{2.3in}{\raggedright \url{https://github.com/zvar-astro/ZVAR-Forced-Phot}\strut} &
  \parbox[t]{\dimexpr 7in-1.1in-2.3in-6\tabcolsep\relax}{\raggedright The pipeline used to generate forced photometry on ZTF sources.\strut} \\[3pt]
\parbox[t]{1.1in}{\raggedright ZVAR Period\strut} &
  \parbox[t]{2.3in}{\raggedright \url{https://github.com/zvar-astro/ZVAR-Period}\strut} &
  \parbox[t]{\dimexpr 7in-1.1in-2.3in-6\tabcolsep\relax}{\raggedright The code used for the ZVAR DR1 periodicity analysis.\strut} \\[3pt]
\parbox[t]{1.1in}{\raggedright ZTF Variability\strut} &
  \parbox[t]{2.3in}{\raggedright \url{https://github.com/zvar-astro/ZVAR-Forced-Phot/metric_suite.py}\strut} &
  \parbox[t]{\dimexpr 7in-1.1in-2.3in-6\tabcolsep\relax}{\raggedright The code used for the ZVAR DR1 variability metrics.\strut} \\[3pt]
\hline
\end{tabular}
\end{table*}

Generative AI has been used to create Figure~\ref{fig:coverage} and build the code to generate the white dwarf results in Section~\ref{sec:results}.
It was also used to generate the numbers in Table \ref{tab:accounting} and  help generate and format Table \ref{tab:source-catalog}.

\section{Data Availability}
The ZVAR DR1 Source Catalog is available at zenodo:...
The total size is 35 GB. 

We request that studies making use of ZVAR data include the following acknowledgement: "This work made use of the ZVAR DR1 catalog and associated light curves, derived from observations obtained by the Zwicky Transient Facility (ZTF) with the Samuel Oschin Telescope 48-inch at the Palomar Observatory. ZVAR is supported by the National Science Foundation (NSF) under Grant No. AST-2509891. ZTF has been supported by NSF Grants No. AST-1440341, AST-2034437, AST-2407588 and the ZTF Collaboration. ZTF operations are conducted by COO, IPAC, and UW."

\section{Acknowledgments}\label{sec:acknowledgements}
MJG, TP, and AJD acknowledge support from the US National Science Foundation through grant AST-2509891. TP and SW acknowledge support from NASA LISA Preparatory Science Grant 80NSSC24K0361. SW acknowledges support from the NASA FINESST Graduate Student Research grant 80NSSC23K1434. M.W.C. acknowledges support from the NSF with grant numbers PHY-2117997, PHY-2409481, and OAC-2609004.

We acknowledge the assistance of April Luce and Hazel Yun Sol Bin in this project.

We acknowledge the valuable contributions of the ZVAR Science Team to the ZVAR Project.  The members include: 
Kevin Burdge (MIT),
Michael Coughlin (University of Minnesota),
Ilaria Caiazzo and Jan van Roestel (Institute of Science and Technology, Austria), 
Maria Drout (University of Toronto, Canada), 
Douglas Finkbeiner (Harvard),
Shravan Hanasoge (Tata Institute for Fundamental Research, India), 
Rocio Kiman (UC Santa Barbara and Caltech),
Yuxi Lu (Ohio State and Columbia University),
Rebecca Phillipson and Scott Engle (Villanova University),
Jean Solmalwar (UC Berkeley and Caltech),  
Paula Szkody (University of Washington), 
and Michele Vallisneri (ETH, Switzerland).

We acknowledge the many contributions of Prof. Shri Kulkarni to the ZTF project, without which ZVAR would never have been possible.

Based on observations obtained with the Samuel Oschin Telescope 48-inch at the Palomar Observatory as part of the Zwicky Transient Facility project. ZTF is supported by the National Science Foundation under Grants No. AST-1440341, AST-2034437, and currently Award AST-2407588 and a collaboration including current partners Caltech, USA; Caltech/IPAC, USA; University of Maryland, USA; University of California, Berkeley, USA; Cornell University, USA; Drexel University, USA; University of North Carolina at Chapel Hill, USA; Institute of Science and Technology, Austria; National Central University, Taiwan, German Center for Astrophysics, Germany, and OKC, University of Stockholm, Sweden. Operations are conducted by Caltech's Optical Observatory, Caltech/IPAC, and the University of Washington at Seattle, USA.

\begin{contribution}
MJG was responsible for writing and submitting the manuscript. MJG and TP are the PIs of ZVAR and obtained the funding for the project. TP also edited the manuscript.
SW carried out the periodicity search and wrote the corresponding sections of the manuscript. AJD generated the variability metrics and carried out significant quality analysis on the data products. ZV wrote the forced photometry code used to produce the light curves. PM and KB produced the initial version of the forced photometry catalog. TC and TdL designed and implemented the web app and API. IC, JvR, and MWC helped with construction and testing.

\end{contribution}

\clearpage

\appendix



\section{ZVAR DR1 Source Catalog Schema}
\label{app:source-catalog}
\suppressfloats[t]

\begin{deluxetable}{lll}[!h]
\label{tab:source-catalog}
\tablecaption{Columns of the ZVAR source catalog. Each row is one source in one ZTF
quadrant; deduplicate on \texttt{ps1\_id} for one row per source. Missing values are
empty fields, except \texttt{gaia\_id}, where 0 means no association exists.
The catalog contains all sources in ZVAR DR1 that have at least
one observation in g-, r-, or i-band.\\}
\tablewidth{\textwidth}
\tabletypesize{\small}
\tablehead{
  \colhead{Column} & \colhead{Type} & \colhead{Description}
}
\startdata
\texttt{ps1\_id} & unsigned long (18 dig.)& \parbox[t]{0.62\textwidth}{\raggedright PanSTARRS1 DR1 object ID; primary key of the ZVAR data product. Stored in the HDF5
\texttt{sources} table as \texttt{ps1\_id}. Never null.\strut} \\[3pt]
\texttt{gaia\_id} & signed long (19 dig.) & \parbox[t]{0.62\textwidth}{\raggedright Gaia DR3 source ID, from the official Gaia DR3 $\times$ PS1 DR1 crossmatch. 0 means
no association exists, not unknown; about 58\% of rows, as PS1 reaches some 2\,mag
deeper than Gaia.\strut} \\[3pt]
\texttt{gaia\_n\_neighbours} & int & \parbox[t]{0.62\textwidth}{\raggedright PS1 sources that are neighbours of this Gaia source. Greater than 1 means the Gaia
source had several PS1 candidates; the best is reported. 0 when \texttt{gaia\_id}
is 0.\strut} \\[3pt]
\texttt{gaia\_n\_mates} & int & \parbox[t]{0.62\textwidth}{\raggedright Other Gaia sources sharing this PS1 source as their best neighbour. Greater than 0
indicates a blend. 0 when \texttt{gaia\_id} is 0.\strut} \\[3pt]
\texttt{ra} & float (deg) & \parbox[t]{0.62\textwidth}{\raggedright Right ascension, ICRS. Inherited from PanSTARRS, not measured by ZTF.\strut} \\[3pt]
\texttt{dec} & float (deg) & \parbox[t]{0.62\textwidth}{\raggedright Declination, ICRS. Inherited from PanSTARRS, not measured by ZTF.\strut} \\[3pt]
\texttt{field} & int & \parbox[t]{0.62\textwidth}{\raggedright ZTF field number.\tablenotemark{a} Values between 245 and 881 are the primary grid,
above 1000 would be the secondary grid (not included in this release).\strut} \\[3pt]
\texttt{ccd} & int, 1--16 & \parbox[t]{0.62\textwidth}{\raggedright ZTF CCD number.\tablenotemark{a}\strut} \\[3pt]
\texttt{quad} & int, 1--4 & \parbox[t]{0.62\textwidth}{\raggedright Quadrant within the CCD.\tablenotemark{a}\strut} \\[3pt]
\texttt{n\_r} & int (epochs) & \parbox[t]{0.62\textwidth}{\raggedright Number of r-band image-difference forced-photometry measurements.\strut} \\[3pt]
\texttt{n\_g} & int (epochs) & \parbox[t]{0.62\textwidth}{\raggedright As above, g band.\strut} \\[3pt]
\texttt{n\_i} & int (epochs) & \parbox[t]{0.62\textwidth}{\raggedright As above, i band; 0 if the quadrant has no i-band matchfile or the source is absent
from it.\strut} \\[3pt]
\texttt{ps1\_g} & float (mag, AB) & \parbox[t]{0.62\textwidth}{\raggedright PanSTARRS1 mean PSF magnitude, g band. Empty where PS1 has no measurement.\strut} \\[3pt]
\texttt{ps1\_r} & float (mag, AB) & \parbox[t]{0.62\textwidth}{\raggedright PanSTARRS1 mean PSF magnitude, r band. Empty where PS1 has no measurement.\strut} \\[3pt]
\texttt{ps1\_i} & float (mag, AB) & \parbox[t]{0.62\textwidth}{\raggedright PanSTARRS1 mean PSF magnitude, i band. Empty where PS1 has no measurement.\strut} \\
\enddata
\tablenotetext{a}{ZTF field numbering is defined in \url{https://github.com/ZwickyTransientFacility/ztf_information}. CCD and readout-quadrant numbering are defined in the
ZTF Science Data System (ZSDS) Explanatory Supplement, IRSA/IPAC:
\url{https://irsa.ipac.caltech.edu/data/ZTF/docs/ztf_explanatory_supplement.pdf}.}
\end{deluxetable}

\clearpage

\section{ZVAR DR1 Excised Period Ranges}
\label{app:excise}

\begin{table}[h!]
\centering
\caption{Excision filter central frequencies, $f_0$, and half-widths, $\Delta f$. 
The reference frequencies are: $f_d = 1.0027372/day$, $f_{lunar} = 0.036601/day$, and $f_{yr} = 0.0027378/day$.}
\begin{tabular}{lcc@{\qquad}lcc}
\hline
\multicolumn{3}{c}{Day frequencies} &
\multicolumn{3}{c}{Year and lunar frequencies} \\
\hline
Description & $f_0$ & $\Delta f$(day$^{-1}$) &
Description & $f_0$ & $\Delta f$(day$^{-1}$)\\
\hline
1 day    & $f_\mathrm{d}$      & $5.00 \times 10^{-2}$ &
1 year   & $f_\mathrm{yr}$     & $2.50 \times 10^{-4}$ \\

2 day    & $f_\mathrm{d}/2$    & $2.00 \times 10^{-2}$ &
2 year   & $f_\mathrm{yr}/2$   & $1.25 \times 10^{-4}$ \\

3 day    & $f_\mathrm{d}/3$    & $1.00 \times 10^{-2}$ &
3 year   & $f_\mathrm{yr}/3$   & $8.35 \times 10^{-5}$ \\

4 day    & $f_\mathrm{d}/4$    & $1.00 \times 10^{-2}$ &
4 year   & $f_\mathrm{yr}/4$   & $6.25 \times 10^{-5}$ \\

5 day    & $f_\mathrm{d}/5$    & $1.00 \times 10^{-2}$ &
5 year   & $f_\mathrm{yr}/5$   & $5.00 \times 10^{-5}$ \\

1/2 day  & $2f_\mathrm{d}$     & $5.00 \times 10^{-2}$ &
1/2 year & $2f_\mathrm{yr}$    & $2.50 \times 10^{-4}$ \\

1/3 day  & $3f_\mathrm{d}$     & $5.00 \times 10^{-2}$ &
1/3 year & $3f_\mathrm{yr}$    & $2.50 \times 10^{-4}$ \\

3/2 day  & $2f_\mathrm{d}/3$   & $2.00 \times 10^{-2}$ &
1/4 year & $4f_\mathrm{yr}$    & $2.50 \times 10^{-4}$ \\

2/3 day  & $3f_\mathrm{d}/2$   & $2.00 \times 10^{-2}$ &
Lunar month & $f_\mathrm{lunar}$ & $5.00 \times 10^{-3}$ \\

3/4 day  & $4f_\mathrm{d}/3$   & $2.00 \times 10^{-2}$ &
& & \\

2/5 day  & $5f_\mathrm{d}/2$   & $2.00 \times 10^{-2}$ &
& & \\

2/7 day  & $7f_\mathrm{d}/2$   & $2.00 \times 10^{-2}$ &
& & \\

1/4 day  & $4f_\mathrm{d}$     & $5.00 \times 10^{-2}$ &
& & \\

1/5 day  & $5f_\mathrm{d}$     & $5.00 \times 10^{-2}$ &
& & \\

1/6 day  & $6f_\mathrm{d}$     & $5.00 \times 10^{-2}$ &
& & \\

1/7 day  & $7f_\mathrm{d}$     & $5.00 \times 10^{-2}$ &
& & \\

1/8 day  & $8f_\mathrm{d}$     & $5.00 \times 10^{-2}$ &
& & \\

1/9 day  & $9f_\mathrm{d}$     & $5.00 \times 10^{-2}$ &
& & \\

1/10 day & $10f_\mathrm{d}$    & $5.00 \times 10^{-2}$ &
& & \\
\hline
\end{tabular}
\label{tab:excision_filters}
\end{table}

\

\bibliographystyle{aasjournal}
\bibliography{refs,refs2}
\label{lastpage}
\end{document}